# FORMATION OF ELECTRIC CHARGES IN MELTING LAYER

A.V. Kochin

Central Aerological Observatory, Dolgoprudny, Moscow Region, 141700 RUSSIA

## 1. INTRODUCTION

The mechanisms for generation of electric charges in clouds have been intensively studied in many countries for a long time, which is an indication of both scientific and applied importance of the subject. The majority of theoretical models describe processes in cumulonimbus clouds. Models dealing with nimbostratus clouds are few and more of a qualitative nature. They do not offer any reliable quantitative estimates of the observed effects.

However, the electric charges are being generated not only in Cb but also in Ns. Moreover, about 80% of the electric discharges to the aircraft occur in Ns clouds (Brylev et al., 1989). This is not an indication of a large electric activity of Ns clouds but, rather is a result of the fact that any flying in vicinity of the cumulonimbus clouds is prohibited because of strong vertical drafts inside them.

Thus, development of a model to explain electrification of the Ns clouds is viewed as an urgent task. Besides, one can reasonably assume that, similar to precipitation development in Ns and Cb, the generation of the electric charges in different types of clouds is based on some common principles, but also has specific features resulting from peculiarities of the given cloud type. Hence, the model should be also valid for explanation of the processes in cumulonimbus clouds.

## 2. ORIGIN OF ELECTRIC FIELD IN CLOUDS

Any model intended for description of electric charge formation and origin of the electric field in the clouds should explain two phenomena, which could be called micro- and macroscale partitioning of the charges.

At the stage of the charge microscale partitioning, the electrically neutral hydrometeors (droplets, snow flakes, hailstones) are charged by elementary electrification mechanisms. For example, a rapid freezing of a droplet is accompanied by breaking small pieces off the ice cover, which could be electrically charged. In response, the initially neutral droplet acquires an equal charge of the opposite sign. Similarly, the air current overflowing a melting graupel pellet, blows off small charged droplets thus charging the graupel pellet with a charge of the opposite sign. A large number of electrification mechanisms have been examined (Mason, 1971; Muchnik, 1974).

However, no significant electric fields are induced because the carriers of charges of different signs are close to each other and mutually compensate for the induced electric fields.

After the microscale partitioning is completed, the charges of different signs should gather in relatively small and spatially separated areas, with the concentration of charges of a given sign exceeding that of the opposite sign (i.e., a macroscale charge partitioning).

## 3. A PHENOMENOLOGICAL DESCRIPTION OF THE PROPOSED MODEL

This paper suggests a mechanism for generation of electric charges within clouds. According to this mechanism
- the microscale partitioning of electric charges takes place within the low melting layer due to breaking of large droplets emerging from melting ice particles,
- the macroscale partitioning is induced by different rates of gravitational sedimentation of the different charge carriers,
- the emerging electric field increases the charge generation rate in the course of the microscale partitioning.

### 3.1 Large droplet formation and breaking

Precipitation particles form from the water vapour mainly in the cold part of the cloud. Then, the ice particles (snow flakes, snow aggregates, graupel pellet) fall down into warm area and turn into droplets thus forming a so called melting

Corresponding author's address: A.V. Kochin, CAO, Pervomaiskaya 3, Dolgoprudny, Moskow region, 141700 Russia, E-mail: alexkochin@mtu-net.ru

layer. Its vertical extension is several hundred metres.

A distance covered by a falling ice particle before its total melting is approximately proportional to a cubic root of the particle mass (Kochin, 1994). In the low melting layer the largest ice particles will be transformed into largest droplets. Drops which diameter exceeds 4 mm are unstable and tend to break into a large number of smaller droplets. The process is as follows. A droplet with an initial diameter of 4-5 mm grows to 40-50 mm, transforming its shape to that of parachute. Then the parachute top breaks into a large number of small droplets, while its bottom yields a few large ones (Mason, 1971). If the ice particle spectrum contains crystals, which produce drops exceeding 4 mm in diameter, these drops will break. The radar observations have proved droplet breaking in the low melting layer in Ns and Cb (Kochin, 1994).

### 3.2 Electric charge generation

A rain droplet in the electric field behaves as a conducting sphere and polarises in response to electric field strength. This is why breaking of a large droplet in the electric field produces electrically charged fragments (Mason, 1971). The upper part of the break droplet transforms into small droplets with electric charges, which sign corresponds to that of the vertical component of electric field. The lower part of the droplet transforms into large droplets with charges of the opposite sign.

Usually the vertical component of the Earth's electric field is negative, so the small and large drops will be charged negatively and positively, respectively. The values of the forming charges are proportional to the field strength (Mason, 1971). Thus, a microscale partitioning of charges takes place.

### 3.3 Macroscale partitioning

Macroscale partitioning of the charges occurs due to different falling velocities of the carriers of charges of different signs. Since the velocities of small droplets are smaller, their concentration in the vicinity of the melting layer will be much greater than concentration of the large droplets. Thus, a negatively charged region forms there.

As the distance from the melting layer increases, the concentration of small droplets declines. This results in a relative growth of large droplet concentration and inception of a positively charged region at some distance below the melting layer.

### 3.4 Enhancement of charge generation

Initial large droplets break in the Earth's electric field (~150 V/m) and then the drop fragments (small and large droplets) leave the layer where the drops are breaking. After that, the electric field in this layer is controlled by interaction of negatively charged small droplets and positively charged large droplets. The small droplets induce a field which sign coincides with that of the Earth's field; a contribution of small droplets into the resulting field will be greater since they are closer to the droplet breaking layer. Hence, the electric field strength in that layer grows up. The value of the breaking charge is proportional to the field strength. Thus, a positive feedback arises which will lead to a continuos growth of the field strength because each subsequent portion of droplets breaks in a stronger electric field and produces a larger charge than the previous one.

## 4. QUANTITATIVE ESTIMATES OF ELECTRIC FIELD STRENGTH GROWTH

The above description of a charge generation mechanism presents only the basics of the proposed model. In order to give quantitative estimates one needs to derive an equation describing the charge generation process with an account of the affecting factors, i.e. vertical and horizontal extension of the droplet breaking layer, ice particle size distribution, vertical draft velocity within the melting layer, etc.

This equation was derived and numerical simulations of electric field strength growth were made by using of the vertical draft velocity and precipitation rate as variables. Ice particle distribution was taking in accordance with Rogers (1976).

The length of the path of a particle before melting is expressed as a linear function in accordance with Kochin (1994)

$$L(d) = 100 d \quad (1)$$

where d and L(d) are given in mm and m, respectively. The horizontal dimention of the droplet breaking layer was taking 300 m in accordance with radar data.

## 5. RESULTS OF NUMERICAL SIMULATIONS

The numerical simulations show that in the course of time the vertical profile of the field strength assumes a complicated shape of varying sign, caused by the vertical extension of the droplet breaking layer. In general, the maximum field strength grows up as precipitation enhances. Depending on the vertical draft velocity the maximum field strength in the melting layer can either be a monotonous function of time or have an oscillating nature, with the charge generation rate reaching maximum in the downdrafts with velocities of 0.5 - 1 m/s.

Temporal variation of the maximum electric field strength $E_{max}$ in a downdraft with a velocity of 0.5 - 1 m/s is perfectly described by (for I>5mm/h, t>60s)

$$E_{max} = 250 \exp\left[0.01\frac{I^2-20}{I^2}t\right] \quad (2)$$

where $E_{max}$ is expressed in V/m, I is precipitation rate, mm/h, t is time, sec.

The results of numerical simulations have also shown that

i) within updrafts or weak (less than 0.3 m/s) downdrafts the electric field strength is about 300 - 3000 V/m, with the maximum values only slightly depending on precipitation rate and having a spatial and temporal variability of charge sign;

ii) maximum electric field strength is attained in a layer with temperature +2°C;

iii) when the downdraft velocity is 0.5 - 1 m/s the charge generation rate is the largest and electric field strength growth rate increases as precipitation enhances. In this case the electric field strength reaches values of $10^5$- $10^6$ V/m after 15 min of raining at a rate of 10-15 mm/h.

## 6. COMPARISON WITH EXPERIMENTAL DATA

The experimental data on electrical activity of the Ns clouds are in good agreement with the theoretical results:

- typical electrical field strength in Ns clouds coincides with the model results according to item i) (Brylev et al., 1989);
- maximum frequency of lightening strokes to aircraft is observed at a level with temperature +1°C (Brylev et al., 1989) ;
- according to estimates by different investigators the electric discharging begins when electric field strength reaches $10^5$ - $10^6$ V/m (Mason, 1971) which requires precipitation rate at least 10 mm/h for more than 15 min.

## 7. ON FEASIBILITY OF PROPOSED MECHANISM IN CB CLOUDS

Apart from others, the proposed mechanism is likely to contribute to electrification of Cb clouds as well. This belief is supported by the fact that the model also agrees with the experimental data on the electrical activity of the Cb clouds.

According to the model results, under a near zero updraft velocity the electric field strength attains a value of 300 - 3000 V/m with no further build-up of the electric field strength. In strong updrafts the electric field strength varies within a range -300 - +300 V/m. This corresponds to the observed values in the non-lightening convective clouds.

In order that the field strength, typical of the thunderstorm, is achieved, it is essential that the precipitation area coincides with the downdrafts.

Usually, an updraft concentrates in the centre of the thunderstorm while downdrafts occupy the cloud periphery. The lower part of the cloud converges the horizontal air currents as the upper part is a region of divergence. In accordance with the proposed model, the charges are to generate on the cloud's periphery. The area of charge generation, comprising individual cells, forms a narrow ring or a sickle. This phenomenon has been already discovered from the experiments (Williams, 1989) but failed to be explained theoretically.

Besides that, the following phenomena should be observed near the melting layer:

- a fast change of value and sign of the hydrometeor's charge;
- maximum values of charges on hydrometeors .

Both effects have been found during aircraft spiral descents on the periphery of thunderstorm (McCready and Proudfit, 1965).

Two-charge clouds are most frequent to occur, with positively charged upper parts and negatively charged middles (Mason, 1971). So, one may believe that positive charges are descending, captured by updraft and carried to the cloud top. A similar process is observed in Cb clouds during the hailstone growth, namely, the large rain droplets are carried to the cloud top. Studies of the hail clouds have shown (Muchnik, 1974) that the hailstone growth is accompanied by a strong radio emission, typical of the thunderstorm inception and development stages, a fact that supports this belief.

In 10-15 min from start of precipitation, the positive charges, captured by updraft, will reach the cloud top. By that time the field strength will build-up to $10^5$ - $10^6$ V/m, resulting in lightening activity development.

In case of a sharp enhancement of downdrafts in a precipitating cloud the change of electric field sign within the melting layer is possible. The reason is that at the early stage of development the field sign is oscillating, so an increase in charge generation rate can "fix" the sign of field, which existed at the moment of the downdraft enhancement. As a result, an opposite (as compared to described above) distribution of cloud charges emerges, something which has been proved by observation (Mason, 1971).

## 8. OUTLOOK FOR FURTHER STUDIES

The described model, however, fails to account for several factors, which strongly affect the charge generation rate.

There is a process, similar to droplet breaking, taking place in the melting layer and capable of generating charges. The thing is that the peripheral branches of melting snow crystals subject to breaking off. If this process is taking place in the electric field, the charged fragments emerge, similar to charge partitioning during droplet breaking. However, this process has not been studied in detail and there are only preliminary theoretical estimates of its effectiveness. Based on these results, a quantitative estimation of electric field growth rate has been made from the model equation (5). It has been found that under exponential ice particle size distribution the crystal destruction provides a fast growth of the electric field up to $10^3$ V/m and its further oscillations.

If the ice particle size spectrum is transformed to a monodispersal one, the described process can lead to field strength values of $10^5$ - $10^6$ V/m even in a light precipitation (about 1-3 mm/h). Another possibility is charge generation in precipitating winter-time warm frontal clouds. This situation can provoke formation of a thin layer with positive temperatures, resulting in the same effect as produced by a transformation of the particle size spectrum to a monodispersal one.

Besides, as soon as the electric field strength reaches $10^4$ - $10^5$ V/m, the smaller droplets become unstable and begin breaking, thus giving birth to additional charge carriers.

As the electric field strength exceeds $10^5$ V/m, another factor comes into play, i.e. the electrostatic interaction becomes comparable with the aerodynamical drag effect. This can alter the falling velocity of charged droplets.

A detail analysis of these processes is the next stage in the proposed model improvement.

## 9. CONCLUSION

In conclusion, one has to mention that despite a satisfactory agreement between obtained theoretical results and the experimental data, special experiments are needed for model verification. The experiment should include registration of starting moment and rate of the droplet breaking (e.g., with radar methods) and measurements of electric field strength. Such an experiment will allow to conclude whether physical bases for the proposed model are correct.

## 10. ACKNOWLEDGEMENT

The author is grateful to I. L. Bushbinder and V. L. Kuznetsov for useful discussions and also to V.R. Megalinsky and colleagues from the Central Aerological Observatory for their assistance in conducting the experiments.